\documentclass[aps,showpacs,superscriptaddress,floatfix]{revtex4}

\usepackage{natbib}
\usepackage{epsf}
\usepackage{bm}
\usepackage{bbm}
\usepackage{dsfont}
\usepackage{graphicx}
\usepackage{dcolumn}
\usepackage{amssymb}
\usepackage{amsmath}
\usepackage{nicefrac}

\newcommand{\addrHeidelberg}{Max--Planck--Institut f\"ur Kernphysik, 
Saupfercheckweg 1, 69117 Heidelberg, Germany}
\newcommand{\addrFreiburg}{Theoretische Quantendynamik, 
Physikalisches Institut, 
Universit\"{a}t Freiburg, 79104 Freiburg im Breisgau, Germany}
\newcommand{\addrParis}{Laboratoire Kastler Brossel, 
\'Ecole Normale Sup\'erieure et
Universit\'{e} Pierre et Marie Curie, Case 74,
75005 Paris, France}
\newcommand{\addrGaithersburg}{National Institute of Standards and Technology,
Mail Stop 8401, Gaithersburg, MD 20899-8401, USA}

\begin{document}

\bibliographystyle{myprsty}

\title{Relativistic and Radiative Energy Shifts for Rydberg States}

\author{Ulrich D. Jentschura}
\affiliation{\addrHeidelberg}
\affiliation{\addrFreiburg}

\author{Eric-Olivier Le~Bigot}
\affiliation{\addrParis}

\author{J\"{o}rg Evers}
\affiliation{\addrFreiburg}

\author{Peter J. Mohr}
\affiliation{\addrGaithersburg}

\author{Christoph H. Keitel}
\affiliation{\addrHeidelberg}
\affiliation{\addrFreiburg}

\begin{abstract} 
We investigate relativistic and quantum electrodynamic effects for
highly-excited bound states in hydrogenlike systems (Rydberg states).  In
particular, hydrogenic one-loop Bethe logarithms are calculated for all
circular states ($l = n-1$) in the range $20 \leq n \leq 60$ and
successfully compared to an existing asymptotic expansion for large
principal quantum number~$n$. We provide accurate expansions of the Bethe
logarithm for large values of~$n$, for $S$, $P$ and circular Rydberg
states. These three expansions are expected to give any Bethe logarithms
for principal quantum number $n > 20$ to an accuracy of five to seven
decimal digits, within the specified manifolds of atomic states. Within
the numerical accuracy, the results constitute unified, general formulas
for quantum electrodynamic corrections whose validity is not restricted to
a single atomic state.  The results are relevant for accurate predictions
of radiative shifts of Rydberg states and for the description of the
recently investigated laser-dressed Lamb shift, which is observable in a
strong coherent-wave light field.
\end{abstract}


\maketitle

%
%
\section{Introduction}
\label{intro}

Circular Rydberg states ($n-1 = l = |m|$) have attracted 
attention in the past two decades, in part 
because the transitions among these states are resonant 
with typical modes of microwave cavities.
A convenient mechanism for the preparation 
of these states has been described in~\cite{HuKl1983,RoEtAl1990,NuEtAl1993}.
When coupled to the modes of a microwave cavity, the 
metastable Rydberg atoms undergo Rabi oscillations, leading 
to a direct verification of the concept of field 
quantization~\cite{BrEtAl1996}. The coupling of the 
resonant cavity to the Rydberg atom can be used
for the generation of entangled Einstein--Podolsky--Rosen
pairs of atoms~\cite{HaEtAl1997}. 
A further field of interest has been the theoretical simulation
of atomic non-dispersive wave packets, which can be realized 
with the ``assistance'' of a background laser field with a
specific ``Kepler'' frequency
(see, e.g.,~\cite{BBKaEb1994,ZaDeBu1997,ShKaEb1998}). 

\begin{figure}
\parbox{7.2cm}{\includegraphics[width=0.9\linewidth]{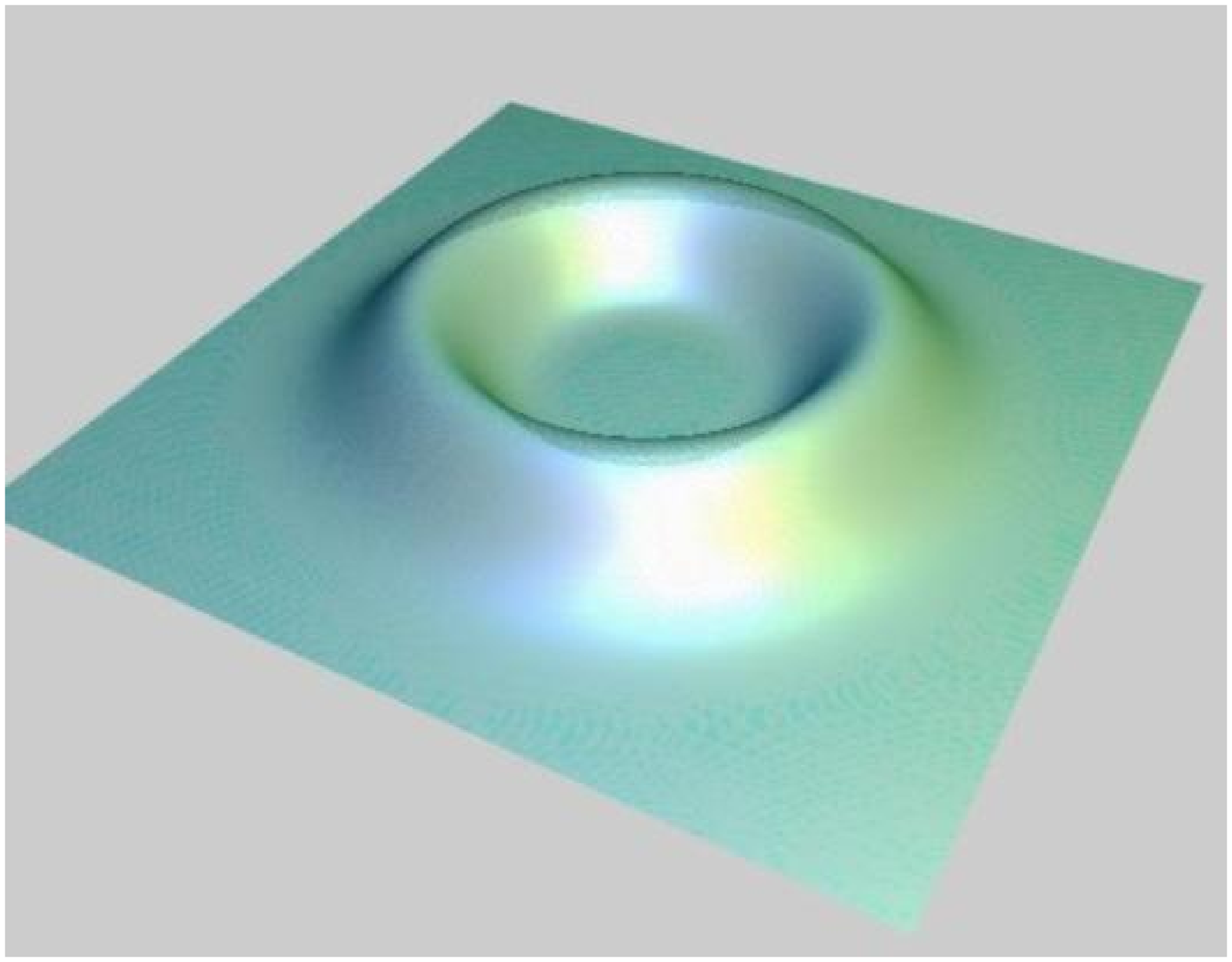}\\ 
\centerline{(a)}}
\parbox{7.2cm}{\includegraphics[width=0.9\linewidth]{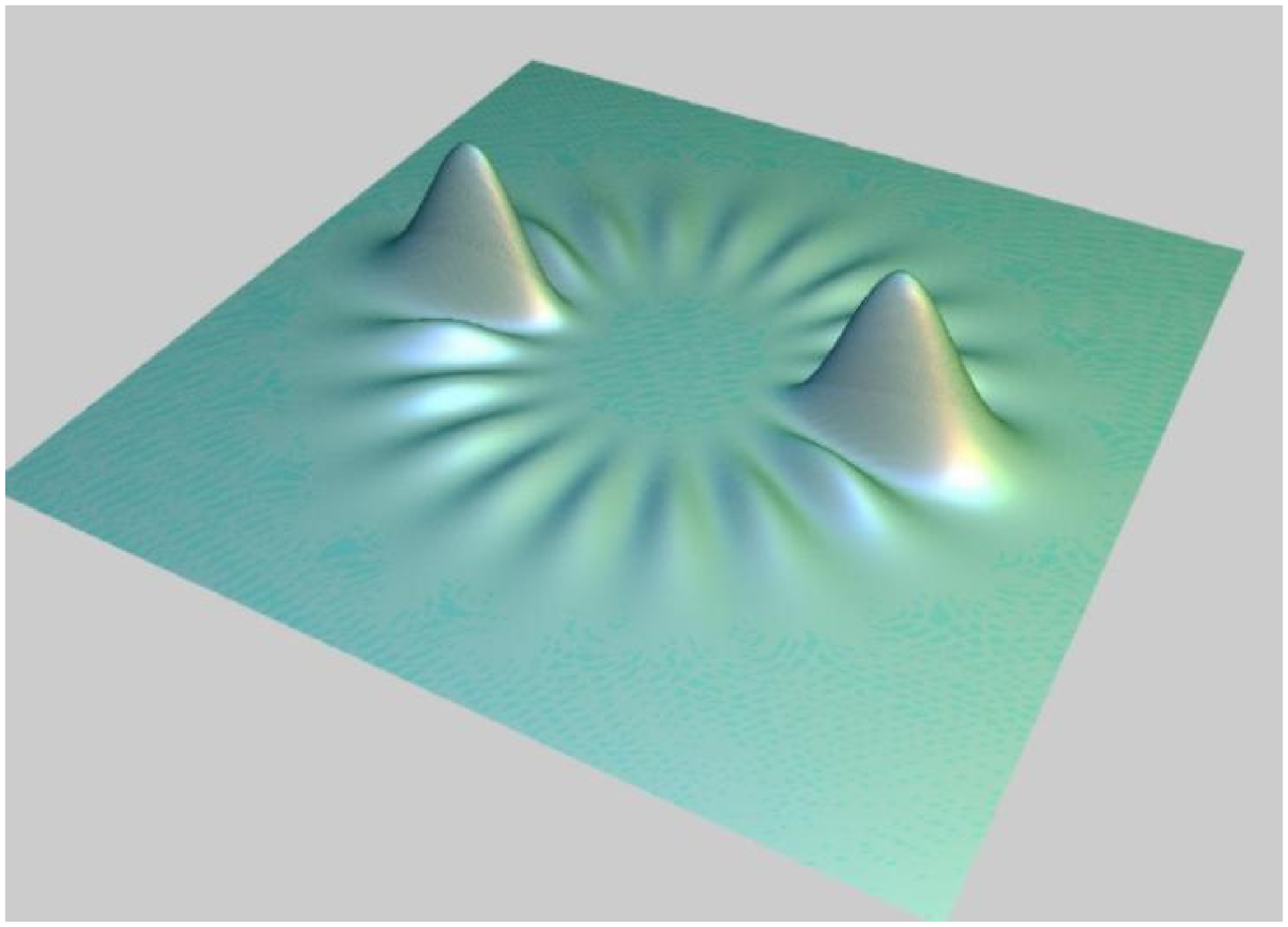}\\ 
\centerline{(b)}}
\caption{\label{fig1} (color online.) 
Plots of the radial
probability density $r^2 |\psi(r,\theta,\phi)|^2$ for 
two (nonrelativistic) states with 
quantum numbers $n=10$ and $l=9$, but different 
magnetic quantum numbers. 
Figure~(a) displays the density of the states with $|m|=9$,
in the plane of constant polar angle $\theta = \pi/2$.
In space, the probability density of the electron in
this maximum-$|m|$ state describes a 
circular shape, which is why
circular states are being viewed as analogues of 
classical planetary motion in the quantum domain.
However, this characteristic circular pattern 
is restricted to the extremal magnetic quantum numbers ($|m|=l$).
Figure~(b) shows the radial 
probability density of the state with $m=0$,
in the plane of constant 
azimuth $\phi = 0$. The two maxima in the 
probability density are in the direction 
of the polar angles $\theta = 0$ and $\theta = \pi$.}
\end{figure}

This partial list of applications is supplemented here
by an investigation of relativistic and radiative effects
in highly-excited 
hydrogenic states, without any consideration of the 
additional interaction with a cavity. 
The electron density of circular Rydberg states ($n-1 = l = \vert m \vert$)
resembles Kepler circular orbits (see Fig.~\ref{fig1}).
However, the characteristic shape of wave function is restricted
to the maximum-$|m|$ subcomponent (see also Fig.~\ref{fig1}).
For states with nonmaximal $|m| < l$, the shape of the electron 
density is manifestly different.
For all circular Rydberg states, 
the wave function is spread out considerably in
comparison to lower states,
on a length scale of $\langle r \rangle \sim 
{\rm const.} \times n^2\,a_{\rm Bohr}$
(see Ref.~\cite{BeSa1957}),
where $r$ is the radial coordinate,
$a_{\rm Bohr}$ is the Bohr radius and $n$ is the
(high) principal quantum number.
Relativistic and radiative effects
in Rydberg atoms allow for a simultaneous perspective
on four different physical regimes:
{\em relativistic} and {\em QED radiative} effects in the transition
in the borderline region
between {\em classical} and {\em quantum} physics.

In addition to the aforementioned reasons for studying Rydberg atoms, the current paper is also inspired
by a recently completed analysis 
of radiative corrections for laser-dressed 
states~\cite{JeEvHaKe2003,JeKe2004aop,EvJeKe2004}.
In~\cite{JeEvHaKe2003,EvJeKe2004}, an 
experimental setup based on the hydrogen $1S$--$2P_j$
($j = \nicefrac{1}{2}, \nicefrac{3}{2}$) 
transition is discussed in detail.
A coherent-wave light source with the appropriate
frequency has recently become available~\cite{EiWaHa2001}. 
Because the $1S$ ground state 
is fully stable against radiative (spontaneous) decay,
the incoherent atomic fluorescence (in the case
of the $1S$--$2P_j$ transition) is described 
to excellent accuracy by the Mollow spectrum~\cite{Mo1969}.
A further promising setup for studying the laser-dressed
Lamb shift would be provided by a transition from 
the metastable $2S$ state to a high-$n$ (Rydberg) $P$ state.
Rydberg states have a small radiative decay width,
which would facilitate the observation of 
the predicted dynamic corrections to the Lamb 
shift~\cite{JeEvHaKe2003,RydDressed}. 
Furthermore, the frequency 
required for the excitation process $2S \Leftrightarrow nP_j$ 
(high $n$) is smaller and therefore {\em a priori} less problematic 
to realize experimentally than the $1S$--$2P_j$
transition. 
In the formalism of~\cite{JeEvHaKe2003,JeKe2004aop,EvJeKe2004},
the self-energy correction analyzed here  contributes
to the Lamb shift of ``bare'' atomic levels; this contribution
is an essential part of
the laser-dressed Lamb shift, which may be 
observed in incoherent light scattering by hydrogenlike atoms.

Another motivation for the 
current work is to find actual 
values of the Bethe logarithms for highly-excited states,
since the Bethe logarithm gives a significant contribution 
to radiative level shifts. This is true of hydrogenlike,
as well as heliumlike~\cite{Ko1999,DrGo1999,Ko2004} and 
lithiumlike~\cite{YaDr2003} systems.
One of the most extensive systematic studies~\cite{DrSw1990} 
of hydrogenic (one-loop) Bethe logarithms currently available in the 
literature extends only up to $n=20$, and only approximate values are
available for higher excited circular Rydberg states~\cite{LBEtAl2003}.
Furthermore, it is of interest to study the asymptotics
of quantum electrodynamic corrections as a function 
of the quantum numbers. Such results can for instance be useful 
in comparing different calculational approaches to these corrections,
and for obtaining general expressions valid for arbitrary quantum 
numbers within a specific manifold of states. 
Certain asymptotic structures have recently been 
conjectured for radiative 
corrections~\cite{JeEtAl2003,Je2003jpa,LBEtAl2003}.

Furthermore, the analysis of highly-excited (bare) atomic states is
motivated by recent experiments: e.g.,
in~\cite{dV2002}, the evaluation of self
energies of circular states of orbital quantum number
$l\simeq 30$ was required. The consideration 
of quantum electrodynamic corrections 
for highly-excited states is also relevant, in part, to the analysis
of quantum electrodynamic corrections 
to dielectronic recombination resonances~\cite{LiEtAl2001}.

In this work, we thus study radiative corrections 
for highly excited hydrogenic states. Specifically,
after recalling basic facts about relativistic
corrections in Sec.~\ref{subsec1}, we proceed to the 
study of radiative effects in Secs.~\ref{subsec2}
and~\ref{subsec3}, with an emphasis on the radiative
decay width of circular states in Sec.~\ref{subsec3}.
The main part of the investigation reported here 
is contained in~\ref{subsec4}, where a numerical calculation 
of Bethe logarithms for highly and very highly excited
circular states is described (see Table~\ref{table1} below). This 
investigation then leads to the asymptotic formulas for 
Bethe logarithms which yield a relative accuracy
of $10^{-5}$ or better, for general states within a
specified manifold. The self-energy corrections
studied in Sec.~\ref{subsec4} are the by far dominant
radiative corrections for circular Rydberg states.
Conclusions are drawn in Sec.~\ref{conclu}.
This paper follows the usual convention for hydrogenic
quantum numbers: the principal quantum number is denoted by~$n$, the
orbital angular momentum by~$l$ and the total electron angular
momentum by ~$j$.  As is customary in the literature, $Z$ denotes the
nuclear charge number of the hydrogenlike ion under consideration,
and $\alpha$ is the fine-structure constant.
Natural units ($\hbar = c = \epsilon_0 = 1$) are
used throughout the text.

%
%
\section{Relativistic and Radiative Effects}
\label{calculation}

%
%
\subsection{Relativistic corrections}
\label{subsec1}

We briefly recall some basic facts about 
relativistic corrections to highly-excited states in hydrogenlike  
systems. 
The numerical values of the coefficients multiplying the 
relativistic corrections of order $(Z\alpha)^4 m c^2$ are typically small
for high-$n$ states. In order to illustrate this point, we briefly recall 
the expansion of the Dirac bound-state energy
in a hydrogenlike system in powers of $Z\alpha$
up to order $(Z\alpha)^4$ (see, e.g.,~\cite[Eq.~(2-87)]{ItZu1980}):
\begin{equation}
\label{enj}
E_{nj} = m - \frac{(Z\alpha)^2\,m}{2 n^2} 
- \frac{(Z\alpha)^4 \, m}{n^3} 
\left[ \frac{1}{2 j + 1} - \frac{3}{8 n} \right] 
+ {\cal O}(Z\alpha)^6\,.
\end{equation}
The $(Z\alpha)^4$-term is the dominant 
relativistic correction for low-$Z$ hydrogenlike 
systems and is responsible, in particular,
for the fine structure. By inspection of Eq.~(\ref{enj}),
we infer that for circular states (which have $j = n - 1 \pm 1/2$), 
the $(Z\alpha)^4$-term scales as $j^{-4} \sim n^{-4}$, which 
contrasts with the familiar $n^{-3}$-scaling valid for fixed~$j$.
As we will see below, \emph{radiative} corrections in Rydberg states also
tend to display faster asymptotic behaviors (in~$n$) than states with
a fixed angular momentum.

%
%
\subsection{Self Energy and Radiative Lifetime}
\label{subsec2}

In this article, we study radiative effects in highly-excited 
hydrogenic energy levels. Our focus is on the 
self-energy and the radiative lifetime.
As pointed out in Ref.~\cite{BaSu1978}, 
the self-energy and the radiative lifetime are
intimately related to each other.
Namely, the radiative lifetime of a hydrogenic state 
is proportional to the imaginary part of the expectation value 
of that state $\langle \psi | \Sigma_{\rm ren} | \psi \rangle$, 
where $\Sigma_{\rm ren}$ is the renormalized self-energy 
operator discussed in Ch.~7 of~\cite{ItZu1980}, and 
$\psi$ is the relativistic (Dirac) wave function.
The self-energy is just the real part of 
$\langle \psi | \Sigma_{\rm ren} | \psi \rangle$ 
and leads to the familiar shift of the 
energy levels. 

As is well known, 
the real part of the one-loop self-energy shift 
$\Delta E_{\rm SE}$ may be written as
\begin{equation}
\label{ESEasF}
\Delta E_{\rm SE} = 
{\rm Re} \langle \psi | \Sigma_{\rm ren} | \psi \rangle =
\frac{\alpha}{\pi} \, \frac{(Z \alpha)^4\,m}{n^3} \, 
F(nl_j,Z\alpha)\,.
\end{equation}
Here, $F(nl_j,Z\alpha)$ is a dimensionless quantity,
and the the notation $nl_j$ follows the usual 
spectroscopic characterization of the hydrogenic state~$\psi$.
It is customary in the literature to suppress
the dependence of $F$ on the quantum numbers $n$, $j$ and $l$
and write $F(Z\alpha)$ for $F(nl_j,Z\alpha)$.

The decay width $\Gamma$ is the reciprocal of the 
lifetime $\tau$ and is given by 
\begin{equation}
\label{defGamma}
\Gamma = \frac{1}{\tau} = -2 \, {\rm Im} \, 
\langle \psi | \Sigma_{\rm ren} | \psi \rangle\,.
\end{equation}
For states with the same angular momentum quantum numbers,
the decay rate decreases with $n$ approximately as $n^{-3}$,
and the lifetime correspondingly goes as $n^3$ 
(see Ref.~\cite{BeSa1957}). This is also 
manifest in the definition~(\ref{ESEasF}) of the scaled self-energy function
$F(nl_j,Z\alpha)$, the structure of which reflects the 
usual, familiar scaling of the effect with $n$
(for fixed $l$ and $j$). 

%
%
\subsection{Radiative Lifetime: Asymptotics for Rydberg States}
\label{subsec3}

To a good approximation, the lifetime of a
highly-excited hydrogenic state does not 
depend on the spin of the electron, and it is only the 
principal quantum number and the orbital angular momentum which enter
into the leading-order (nonrelativistic)
expression for the decay rate of excited 
hydrogenic states. 

We denote the (leading-order) decay width of a state with principal
quantum number~$n$ and orbital angular momentum~$l$
by the symbol $\Gamma(n,l)$. Here, we are particularly interested in 
states with $n = l+1$, and our goal is to 
study the dependence of the lifetime on the
bound-state quantum numbers. For the manifold of states
with $n = l+1$, based on the formalism introduced in~\cite{BaSu1978},
we obtain the following general expression,
\begin{subequations}
\begin{eqnarray}
\label{exactgamma}
\Gamma(n, n-1) &=& \frac{2\alpha}{3} \frac{(Z\alpha)^4 m}{n^5} 
\left\{ \frac{n^{2n+1} (n - 1)^{2n-2}}
{(n - \nicefrac{1}{2})^{4 n - 1}} \right\} \\[2ex]
&=& 
\label{firstterms}
\frac{2\alpha}{3}\, \frac{(Z\alpha)^4 m}{n^5} 
\times \left\{ 1 + 
\frac{1}{n} + 
\frac{7}{8 n^2} + 
\frac{35}{48 n^3} \right. 
\nonumber\\[2ex]
& & \left. + \frac{229}{384 n^4} +
\frac{233}{480 n^5} +
{\cal O}(n^{-6}) \right\}\,.
\end{eqnarray}
\end{subequations}
The asymptotic expansion (\ref{firstterms}) about 
large $n$ works surprisingly well,
even at low principal quantum number. 
For $n=2$, the first few terms listed in 
(\ref{firstterms}) reproduce the exact formula (\ref{exactgamma}) 
to within $5\,\%$ accuracy. For $n=40$, the difference is less than
$10^{-10}$ in relative units. Equation (\ref{firstterms})
implies that the
lifetime of circular Rydberg states with $l = n-1$ scales effectively as 
$n^5$, in contrast to the familiar $n^3$-scaling which is valid
for fixed $l$, but varying $n$.

Based on the well-behaved asymptotic structure of the leading non-relativistic contribution to
${\rm Im}\langle \psi | \Sigma_{\rm ren} | \psi \rangle$
for high quantum numbers, we now turn our 
attention to the real part
${\rm Re} \langle \psi | \Sigma_{\rm ren} | \psi \rangle$.

%
%
\subsection{Self--Energy}
\label{subsec4}

Vacuum-polarization effects are negligible for
high-$l$ circular Rydberg states [the radial 
component of the wave function
scales as $(Z\alpha m r)^l$ for small $r$]. The dominant 
radiative correction to hydrogenic energy levels is given by the (one-photon)  
self energy
\[
\includegraphics*[width=2.5cm]{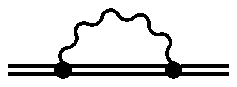},
\]
which is a process in which the bound electron (double line) emits and
re-absorbs a photon (wavy line); this effect
shifts the energies predicted by the Dirac equation.

We now turn to the investigation of the energy 
shift $\Delta E_{\rm SE}$, which is 
defined in Eq.~(\ref{ESEasF}) and corresponds to the above diagram.
The semi-analytic expansion of $F(nl_j,Z\alpha)$
about $Z\alpha = 0$ for a general atomic state with quantum numbers $n$,
$l$ and $j$ gives rise to the expression~\cite{ErYe1965a}
\begin{equation}
\label{defFLO}
F(nl_j,Z\alpha) = 
A_{41}(nl_j) \, \ln[(Z \alpha)^{-2}] + A_{40}(nl_j) + {\cal O}(Z\alpha)\,.
\end{equation}
The $Z\alpha$-expansion is semi-analytic, i.e., it involves powers of
$Z\alpha$ and of~$\ln[(Z\alpha)^{-2}]$. 
The $A$ coefficients have two indices, the first of which denotes the
power of $Z\alpha$ [including those powers contained in
Eq.~(\ref{ESEasF})], while the second index denotes the power of the
logarithm $\ln[(Z \alpha)^{-2}]$.

\begin{figure}[tbh]
\begin{center}
\includegraphics[width=0.6\linewidth]{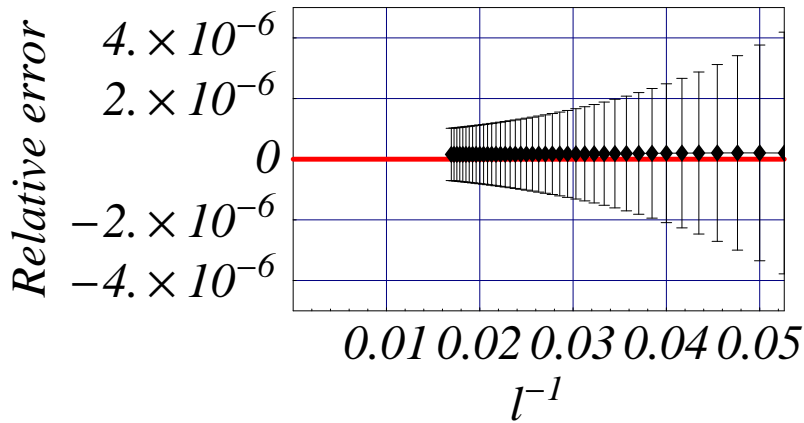}
\end{center}
\caption{\label{fig2} (color online.) Relative difference between the 
Bethe logarithm values presented in Tab.~\ref{table1} and 
the asymptotic formula~(\ref{circasym}), for circular Rydberg states 
($l=n-1$) in the range $20 < n \leq 60$. Since zero is 
contained in all the error bars, the Bethe logarithms in 
Tab.~\ref{table1} are fully consistent with the
approximation~(\ref{circasym}). The error bars in the figure
correspond to the uncertainty in the numerical
coefficients of Eq.~(\ref{circasym}).}
\end{figure}

The coefficient $A_{41}(nl_j)$ assumes a value of $4/3$ for 
$S$~states and vanishes for all non-$S$ states.
The general formula for $A_{40}$ 
reads (see, e.g., \cite{SaYe1990,MoTa2000})
\begin{equation}
\label{A40gen} 
A_{40} = \frac{10}{9}\,\delta_{l0}
-\frac{1-\delta_{l0}}{2\kappa\,(2 l + 1)} - \frac{4}{3}\,\ln k_0(n, l),
\end{equation}
where $\kappa = 2 \, (l-j) \, (j + 1/2)$.
The Bethe logarithm $\ln k_0(n, l)$ is an inherently nonrelativistic
quantity, whose expression in natural units reads
\begin{eqnarray}
\label{bethelog}
\lefteqn{\ln k_0(n, l) = \frac{n^3}{2 (Z\alpha)^4\,m} }                           \\\nonumber
& & \quad \times \left< \phi \left| \frac{p^i}{m} \,
\left( H_{\rm S} - E_n \right) \, 
\ln \left[ \frac{2 \left| H_{\rm S} - E_n \right|}{(Z\alpha)^2\,m} \right] \,
\frac{p^i}{m} \right| \phi \right>\,. 
\end{eqnarray}
Here, $H_{\rm S}$ is the nonrelativistic Coulomb Schr\"odinger
Hamiltonian, $p^i$~is the $i$th Cartesian component
of the momentum operator, $E_n$ and $\phi$ are
respectively the (nonrelativistic) 
energy and the wave-function of a state with the quantum
numbers~$(n,l)$, and the summation over Cartesian coordinates
($i=1,2,3$) is implicit.

Bethe logarithms have been studied 
for all hydrogenic states with $n \leq 20$,
by various authors and with increasing accuracy
due to advances in the algorithms used 
and in the computing technology available~\cite{Be1947,BeBrSt1950,Ha1956,
ScTi1959,Li1968,Hu1969,KlMa1973,DrSw1990,GoDr2000}. 
While the calculational difficulties
(numerical convergence properties of the involved 
hypergeometric functions) increase as the principal quantum number
increases, the evaluation of the hydrogenic nonrelativistic 
Bethe logarithm can be regarded as an easy computational
task from the perspective of our current understanding 
of bound states, and of modern computer technology. 
In the current work, we employ the technique previously 
discussed in~\cite{JePa1996,LBEtAl2003}, where the relevant 
matrix element of the hydrogenic wavefunction~\cite{SwDr1991abc} is expressed 
in terms of hypergeometric functions~${}_2 F_1$, 
which are then suitably integrated. 
For high energies of the virtual photon, we found it useful 
to employ the combined nonlinear-condensation
transformation~\cite{JeMoSoWe1999,AkSaJeBeSoMo2003} in order to
accelerate the convergence of the series representation 
of these hypergeometric functions.

One of the results we have obtained reads
\begin{equation}
\label{specres}
\ln k_0(n=20, l=19)=-0.000~008~084~977~837~087~891(1)\,,
\end{equation}
a value which confirms the result obtained previously
in Ref.~\cite{DrSw1990} for this state.
The error is due to the uncertainty in the final
integration over the virtual photon energy.
While the numerical accuracy of this result 
is merely of academic interest, we found it useful
to verify the accuracy of our computational method 
against the basis-set techniques employed in~\cite{DrSw1990,GoDr2000}.

%
%
\begin{center}
\begin{table}[tbh]
\begin{center}
\begin{minipage}{12cm}
\caption{\label{table1} Values of the Bethe logarithm 
for circular Rydberg states (for which $l = n-1$), in the 
range $20 < n \leq 60$. According to
Eqs.~(\ref{defFLO}) and (\ref{A40gen}), the Bethe 
logarithm $\ln k_0$ contributes significantly to the 
radiative correction to the energy of hydrogenic states.
All explicit results found here agree with a previously published 
expansion~\cite[Eq.~(37)]{LBEtAl2003}, as well as with the new
asymptotic expansion~(\ref{circasym}), as is shown in 
Fig.~\ref{fig2} (within the numerical uncertainties in the 
asymptotic coefficients).}
\begin{tabular}{c@{\hspace*{0.5cm}}r@{\hspace*{1cm}}c@{\hspace*{0.5cm}}r}
\hline
\hline
\rule[-3mm]{0mm}{8mm} $n$ &
 $\ln k_0(n, n-1)$ &
 $n$ &
 $\ln k_0(n,n-1)$ \\
\hline
$21$ & $ -0.69410~65660~642(1) \times 10^{-5} $ & $41$ & $ -0.08782~81749~454(1) \times 10^{-5} $ \\ 
$22$ & $ -0.60031~18198~591(1) \times 10^{-5} $ & $42$ & $ -0.08158~06567~501(1) \times 10^{-5} $ \\ 
$23$ & $ -0.52267~98437~374(1) \times 10^{-5} $ & $43$ & $ -0.07591~18687~638(1) \times 10^{-5} $ \\ 
$24$ & $ -0.45787~44770~701(1) \times 10^{-5} $ & $44$ & $ -0.07075~63248~884(1) \times 10^{-5} $ \\ 
$25$ & $ -0.40335~28104~402(1) \times 10^{-5} $ & $45$ & $ -0.06605~72331~309(1) \times 10^{-5} $ \\ 
$26$ & $ -0.35715~36036~972(1) \times 10^{-5} $ & $46$ & $ -0.06176~51782~987(1) \times 10^{-5} $ \\ 
$27$ & $ -0.31774~78119~539(1) \times 10^{-5} $ & $47$ & $ -0.05783~70279~429(1) \times 10^{-5} $ \\ 
$28$ & $ -0.28393~14416~323(1) \times 10^{-5} $ & $48$ & $ -0.05423~50198~720(1) \times 10^{-5} $ \\ 
$29$ & $ -0.25474~77154~839(1) \times 10^{-5} $ & $49$ & $ -0.05092~59980~293(1) \times 10^{-5} $ \\ 
$30$ & $ -0.22942~98276~714(1) \times 10^{-5} $ & $50$ & $ -0.04788~07701~395(1) \times 10^{-5} $ \\ 
$31$ & $ -0.20735~83589~839(1) \times 10^{-5} $ & $51$ & $ -0.04507~35657~214(1) \times 10^{-5} $ \\ 
$32$ & $ -0.18802~92605~357(1) \times 10^{-5} $ & $52$ & $ -0.04248~15771~656(1) \times 10^{-5} $ \\ 
$33$ & $ -0.17102~95447~205(1) \times 10^{-5} $ & $53$ & $ -0.04008~45698~251(1) \times 10^{-5} $ \\ 
$34$ & $ -0.15601~86575~648(1) \times 10^{-5} $ & $54$ & $ -0.03786~45496~597(1) \times 10^{-5} $ \\ 
$35$ & $ -0.14271~40813~819(1) \times 10^{-5} $ & $55$ & $ -0.03580~54790~517(1) \times 10^{-5} $ \\ 
$36$ & $ -0.13088~01169~928(1) \times 10^{-5} $ & $56$ & $ -0.03389~30330~782(1) \times 10^{-5} $ \\ 
$37$ & $ -0.12031~90769~194(1) \times 10^{-5} $ & $57$ & $ -0.03211~43898~736(1) \times 10^{-5} $ \\ 
$38$ & $ -0.11086~43219~585(1) \times 10^{-5} $ & $58$ & $ -0.03045~80498~130(1) \times 10^{-5} $ \\ 
$39$ & $ -0.10237~47182~006(1) \times 10^{-5} $ & $59$ & $ -0.02891~36791~317(1) \times 10^{-5} $ \\ 
$40$ & $ -0.09473~01966~836(1) \times 10^{-5} $ & $60$ & $ -0.02747~19743~302(1) \times 10^{-5} $ \\ 
\hline
\hline
\end{tabular}
\end{minipage}
\end{center}
\end{table}
\end{center}

In Tab.~\ref{table1}, we present accurate numerical
values for the Bethe
logarithm~(\ref{bethelog}) of circular Rydberg states, for 
$21 \leq n \leq 60$.  
These values are in agreement with the truncated asymptotic
expansion found in Ref.~\cite[Eq.~(37)]{LBEtAl2003}.
This asymptotic expansion had been derived on the 
basis of numerical data reported in 
Ref.~\cite{DrSw1990}, where the range $1 \leq n \leq 20$
had been covered. Based on the numerical data of Tab.~\ref{table1},
which cover a wider range of principal quantum numbers
as compared to Ref.~\cite{DrSw1990},
we obtain the following improved asymptotics
for the Bethe logarithm of circular Rydberg states:
\begin{eqnarray}
\label{circasym}
l^3 \times  \ln k_0(l+1, l) 
&\simeq&
-0.05685281(3) + \frac{0.0248208(6)}{l} 
+ \frac{0.03814(2)}{l^2}
- \frac{0.1145(5)}{l^3} 
+ \frac{0.166(3)}{l^4} 
- \frac{0.22(2)}{l^5} \,,
\end{eqnarray}
where terms of order~$l^{-k}$ with $k \geq 6$ are neglected.
The algorithm described in the Appendix of~\cite{LBEtAl2003}
was used for obtaining the coefficients of this asymptotic expansion.
All the coefficients in (\ref{circasym}) separately are
in very good agreement
with those found previously~\cite[Eq.~(37)]{LBEtAl2003}.
The Bethe logarithm values in
Tab.~\ref{table1} are fully consistent with the above truncated
expansion: they fall inside the error bars of~(\ref{circasym}), as is
illustrated in Fig.~\ref{fig2}.  Furthermore, the ``expectation''
values for the coefficients of formula~(\ref{circasym}) correctly
``predict'' about seven digits of the actual numerical value of the
Bethe logarithm in Tab.~\ref{table1}, as illustrated in
Fig.~\ref{fig2}.  Based on the very good consistency of the asymptotic
expansion with the actual numerical data (Fig.~\ref{fig2}), we conjecture
that the ``expectation'' values of the expansion (\ref{circasym})
give the Bethe logarithm of all circular Rydberg states with $n > 20$ with
a relative precision better than~$10^{-6}$, as can be expected from
Fig.~\ref{fig2}. Furthermore,
we would like to conjecture here that the expansion (\ref{circasym})
not only represents a ``polynomial fit'' to the numerical
data, but that it represents the true asymptotic expansion 
of the Bethe logarithm for high quantum numbers. 

For $S$ states, based on data available for 
principal quantum numbers $n \leq 20$~(see Ref.~\cite{DrSw1990}), 
we obtain the following asymptotic expansion as a function
of the principal quantum number $n$,
\begin{eqnarray}
\label{Sasym}
\ln k_0(n, l=0) 
&\simeq&
2.72265434(5) + \frac{0.000000(5)}{n} 
+ \frac{0.55360(5)}{n^2}
- \frac{0.5993(5)}{n^3} 
+ \frac{0.613(7)}{n^4} 
- \frac{0.60(5)}{n^5} \,,
\end{eqnarray}
where terms of order~$n^{-k}$ with $k \geq 6$ are neglected.  By a
reasoning similar to the one above, we expect the 
expansion~(\ref{Sasym}), with the ``expectation'' values
employed for the coefficients,
to give the Bethe logarithm of all $S$~states with
$n>10$ to better than $2\cdot 10^{-7}$ in relative units.

Finally, for high-$n$ $P$ states, we obtain
\begin{eqnarray}
\label{Pasym}
\ln k_0(n, l=1) 
&\simeq&
-0.0490545(1) + \frac{0.000000(5)}{n} 
+ \frac{0.20530(15)}{n^2}
- \frac{0.599(5)}{n^3} 
+ \frac{1.45(10)}{n^4} 
- \frac{3(1)}{n^5} \,.
\end{eqnarray}
where again terms of order~$n^{-k}$ with $k \geq 6$ are neglected.
We expect Eq.~(\ref{Pasym}) to reproduce
the exact Bethe Logarithms of all $P$~states for $n > 10$ to a 
relative accuracy better than $5 \cdot 10^{-5}$.

%
%
\section{Conclusions}
\label{conclu}

We have studied relativistic and radiative corrections for
highly-excited hydrogenic states.  In addition to $S$ and $P$~states,
we have concentrated on circular Rydberg states (for which $l=n-1$).
We have observed on the relativistic and radiative corrections
considered here that circular Rydberg state corrections have a faster
asymptotic behavior (as a function of~$n$) than states with a fixed
angular momentum.

The 
following scaling properties hold for the dominant relativistic and 
radiative effects on circular hydrogenic Rydberg states:
{\em (i)} The leading relativistic corrections to the energy levels (of order $(Z\alpha)^4 m c^2$) scale
as $n^{-4}$.
{\em (ii)} Radiative decay rates scale as $n^{-5}$
[see Eq.~(\ref{firstterms})].
{\em (iii)} The Bethe logarithm contribution to 
the self-energy shift~(\ref{ESEasF})
scales as $n^{-6}$.
Note the three inverse powers of $l$ in Eq.~(\ref{circasym})
which have to be taken into consideration in addition 
to the three inverse powers of $n$ originating from (\ref{ESEasF}).
By contrast, the Bethe logarithm contribution for highly-excited 
$S$ and $P$~states scales as $n^{-3}$ [see Eqs.~(\ref{Sasym})
and (\ref{Pasym})].
{\em (iv)} The anomalous magnetic moment contribution 
to the Lamb shift [which corresponds to the term 
$-\frac{1}{2\kappa (2 l +1)}$
in Eq.~(\ref{A40gen})] scales as $n^{-5}$ for circular 
Rydberg states and thus dominates over the Bethe logarithm
in the limit of large $l = n-1$.

In this article, we also provide an exact result
for the decay rate of highly-excited hydrogenic
Rydberg states [see Eq.~(\ref{exactgamma})].
Furthermore, we obtain accurate values (see Tab.~\ref{table1}) for
the Bethe logarithms of circular Rydberg states (for which the orbital
angular quantum number $l$ is related to the principal quantum number
as $l=n-1$). These values are in a range of quantum numbers $21 \leq n
\leq 60$, where no such data have been available up to now, to the best
of our knowledge.  These results confirm that the dependence of
radiative corrections on the bound-state quantum numbers can be
represented, to a very good accuracy, by asymptotic expansions about large
$n$ which involve negative integer powers [see
Eqs.~(\ref{firstterms}),~(\ref{circasym}),~(\ref{Sasym})
and~(\ref{Pasym})].  This functional dependence has recently been
observed for a number of QED
effects~\cite{JeEtAl2003,Je2003jpa,LBEtAl2003}.  We conjecture here
that the Bethe logarithm 
expansions~(\ref{circasym}),~(\ref{Sasym}) and~(\ref{Pasym})
represent the true asymptotic expansions about large
quantum numbers (to the precision indicated in the text).

The results obtained here
are relevant for any analysis (e.g.,~\cite{dV2002}) 
where the evaluation of the self
energy of circular ($n=l+1$) states of orbital quantum number
$l > 20$ is required. They are also relevant for the 
description of the ``bare'' Lamb shift contribution 
to the Lamb shift of laser-dressed states~\cite{Mo1969,%
JeEvHaKe2003,JeKe2004aop,EvJeKe2004} in 
experiments with Rydberg states.
A possible experimental setup would involve
the metastable $2S$ state and an excitation to a
high-$n$ $P$ state. The required frequency
is smaller than the one required for the excitation
$1S \Leftrightarrow 2P_j$, which has been studied in
detail in~\cite{EvJeKe2004}. Moreover, both states
involved in the transition $2S \Leftrightarrow nP_j$ 
(high $n$) have a long lifetime, which would 
facilitate the observation of the dressed-state
corrections to the Lamb shift~\cite{JeEvHaKe2003,JeKe2004aop,EvJeKe2004}.
The formalism required for the analysis of 
further corrections to the laser-dressed Lamb shift
has recently been described in~\cite{EvJeKe2004}.
The additional corrections are
due to ionization, counter-rotating terms
(Bloch--Siegert shifts, see~\cite{BlSi1940}) and 
nonresonant levels, as well as laser-field configuration
dependent effects.

Finally, we note that an accurate investigation
of the Lamb shift of the (numerous) different magnetic subcomponents
of circular Rydberg levels would allow for a sensitive test of
the recently proposed, somewhat speculative ``polarized Lamb shift''
(see Ref.~\cite{ChSJTu2001})
that predicts a breaking of the rotational symmetry
imposed on quantum electrodynamics.

\section*{Acknowledgments}

U.D.J.~and E.O.L. acknowledge support from the National Institute
of Standards and Technology during a number of research appointments.
P. Indelicato is acknowledged for helpful discussions.
J.~Evers acknowledges support from the National German Academic 
Foundation.


\begin{thebibliography}{10}
\bibitem{RydDressed}
By contrast, direct transitions among Rydberg states are probably not a
  suitable environment for the study of the dressed Lamb shift in an intense
  laser field: the problem is that a laser which drives a circular Rydberg
  transition between levels with $n,l \Leftrightarrow n+1,l+1$ is approximately
  resonant not only with the desired transitions, but also with
  the transition $n+1,l+1 \Leftrightarrow n+2,l+2$, in the limit of
  large $n$. Therefore, the level scheme used in the analysis of the
  laser-dressed Lamb shift would have to be substantially generalized in order
  to accomodate for the many (near-resonant) energy levels.
\end{thebibliography}
\end{document}